\documentclass[twocolumn]{aastex63}

\usepackage{amsmath}

\shorttitle{RADIAL ELONGATIONS OF X-RAY CAVITIES}
\shortauthors{GUO}

\begin{document}
\bibliographystyle{aasjournal} 
 
\title{Probing the Physics of Mechanical AGN Feedback with Radial Elongations of X-ray Cavities}

\author{Fulai Guo}
\affiliation{Key Laboratory for Research in Galaxies and Cosmology, Shanghai Astronomical Observatory, Chinese Academy of Science, 80 Nandan Road, Shanghai 200030, China; fulai@shao.ac.cn}
\affiliation{University of Chinese Academy of Sciences, 19A Yuquan Road, Beijing 100049, China}

\begin{abstract}

Mechanical active galactic nucleus (AGN) feedback plays a key role in massive galaxies, galaxy groups and clusters. However, the energy content of AGN jets that mediate this feedback process is still far from clear. Here we present a preliminary study of radial elongations $\tau$ of a large sample of X-ray cavities, which are apparently produced by mechanical AGN feedback. All the cavities in our sample are elongated along the angular (type-I) or jet directions (type-II), or nearly circular (type-III). The observed value of $\tau$ roughly decreases as the cavities rise buoyantly, confirming the same trend found in hydrodynamic simulations. For young cavities, both type-I and II cavities exist, and the latter dominates. Assuming a spheroidal cavity shape, we derive an analytical relation between the intrinsic radial elongation $\bar{\tau}$ and the inclination-angle-dependent value of $\tau$, showing that projection effect makes cavities appear more circular, but does not change type-I cavities into type-II ones, or vice versa. We summarize radial elongations of young cavities in simulations, finding that $\bar{\tau}$ increases with the kinetic fraction of AGN jets. While mild jets always produce type-II cavities, thermal-energy-dominated strong jets produce type-I cavities, and kinetic-energy-dominated strong jets produce type-II cavities. Our results suggest that some AGN jets are strong and dominated by thermal energy (or cosmic rays). However, these jets do not dominate in AGN feedback. If most jets are dominated by non-kinetic energies, they should be mainly mild jets. If most jets are strong, they must be mainly dominated by the kinetic energy. 

\end{abstract}

\keywords{
galaxies: active --- galaxies: clusters: general --- galaxies: clusters: intracluster medium --- galaxies:jets --- X-rays: galaxies: clusters
}

\section{Introduction}
\label{section:intro}

 \begin{figure}
   \centering
     \epsscale{0.75}
\includegraphics[width=0.35\textwidth]{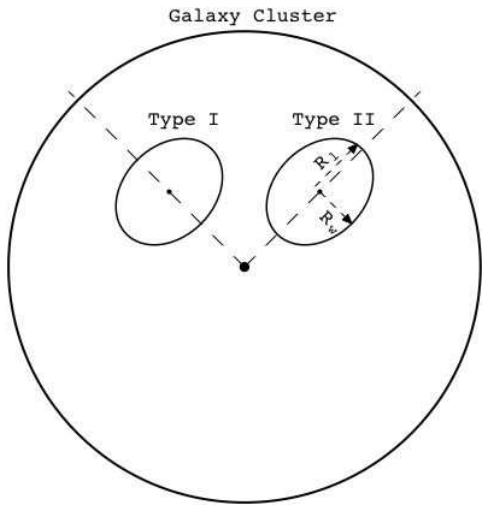} 
\caption{Sketch of type I and II X-ray cavities in a galaxy cluster. $R_{\rm l}$ and $R_{\rm w}$ are the semi axes along the jet direction and the angular direction perpendicular to the jet direction, respectively. The jet direction here is defined as the radial direction from the cluster center to the cavity center. Type I cavities are oblate, elongated along the angular direction, while type II cavities are prolate, elongated along the jet direction. Nearly circular type-III cavities with $R_{\rm l}\approx R_{\rm w}$ have also been found in galaxy clusters. The radial elongation of a cavity is defined as $\tau=R_{\rm l}/R_{\rm w}$.}
 \label{plot1}
 \end{figure} 
 
Mechanical feedback from active galactic nuclei (AGNs) plays a key role in the evolution of massive elliptical galaxies, galaxy groups, and clusters, suppressing cooling flows and the associated star formation activities in central galaxies (\citealt{mcnamara07}; \citealt{mcnamara12}; \citealt{li15}; \citealt{soker16}; \citealt{werner19}). Direct evidence for the operation of AGN feedback comes from mounting detections of ``X-ray cavities" in deep X-ray images of galaxy groups and clusters, apparently evolved from the interaction of AGN jets with the hot intracluster medium (ICM; \citealt{boehringer93}; \citealt{fabian02}; \citealt{birzan04}; \citealt{croston11}, \citealt{vagshette19}). The properties of X-ray cavities may thus contain important information of mechanical AGN feedback.

The enthalpy of X-ray cavities has been widely used to estimate the energetics of mechanical AGN feedback \citep{birzan04,rafferty06,hlavacek12,hlavacek15}. For a cavity with volume $V$, its enthalpy can be written as
   \begin{eqnarray}
H\equiv eV+pV= \frac{\gamma}{\gamma -1}pV {\rm ,}
   \end{eqnarray}
where $e$, $p$, and $\gamma$ are the energy density, the pressure, and the adiabatic index of the plasma inside the cavity, respectively. Under the ``slow-piston" approximation for quasi-static point outbursts in a uniform medium, the total energy of the outburst creating the cavity is $E_{\rm jet}=H$. Assuming that the cavity is mainly filled with relativistic cosmic rays, one has $\gamma=4/3$, $E_{\rm jet}=4pV$, and a low energy coupling efficiency between the outburst and the ambient medium $\eta_{\rm cp} = (E_{\rm jet}-eV)/E_{\rm jet}=(\gamma-1)/\gamma=0.25$ \citep{duan20}. For realistic jet outbursts in galaxy clusters, recent hydrodynamic simulations by \citet{duan20} show that the energy coupling efficiency is much higher with $\eta_{\rm cp} \approx 0.7$-$0.9$, and according to the definition of $\eta_{\rm cp}$, $E_{\rm jet}=eV+\eta_{\rm cp}E_{\rm jet}$, this 
leads to significantly higher estimates for the jet energy $E_{\rm jet}=pV/[(1-\eta_{\rm cp})(\gamma-1)] \approx 10$-$30pV$. 

In addition to the cavity volume, its shape may also contain important information about mechanical AGN feedback. Hydrodynamic jet simulations by \citet{guo15} and \citet{guo16} suggest that the shape of young X-ray cavities recently created by the jet-ICM interaction can be used to probe jet properties, while the shape of old X-ray cavities is affected by the level of viscosity in the ICM. Kinetic-energy-dominated jets on kpc scales typically produce young X-ray cavities more elongated along the jet direction than non-kinetic-energy-dominated jets, which may be energetically dominated by thermal energy, cosmic rays, or magnetic fields. Hydrodynamic simulations of mechanical AGN feedback often adopt kinetic-energy-dominated jets (e.g., \citealt{gaspari11}; \citealt{yang16}; \citealt{guo18}; \citealt{martizzi19}; \citealt{bambic19}), while recent AGN feedback simulations also start to investigate cosmic-ray-dominated jets (\citealt{guo11}; \citealt{ruszkowski17}; \citealt{yang19}; \citealt{wang20}). \citet{duan20} recently found that strong non-kinetic-energy-dominated jets are much more effective in delaying the onset of cooling catastrophe than kinetic-energy-dominated jets with the same power. The particle content in AGN jets and X-ray cavities has also been investigated observationally (e.g., \citealt{croston08}; \citealt{birzan08}; \citealt{croston14}).

In this paper, we extend our previous theoretical studies in \citet{guo15} and \citet{guo16} on the cavity shape, and present a preliminary study on the shape of observed X-ray cavities, focusing on their radial elongations. As a ``zeroth-order" approximation, observed X-ray cavities are often approximated as ellipses, and as seen in Section 2, observed X-ray cavities are usually elongated along either the jet direction or the angular direction, which is perpendicular to the jet direction (see also \citealt{birzan04,rafferty06,hlavacek12}). Here the jet direction is defined as the radial direction from the cluster center to the cavity center. In Figure 1, we show a sketch of these two types of X-ray cavities. The radial elongation of a given X-ray cavity may be defined as $\tau=R_{\rm l}/R_{\rm w}$, where $R_{\rm l}$ is the semi axis along the jet direction and $R_{\rm w}$ is the semi axis along the angular direction. In this paper, we refer to cavities with $\tau < 1$ as type I cavities and those with $\tau > 1$ as type II cavities. Nearly circular cavities with $R_{\rm l}\approx R_{\rm w}$ have also been found in galaxy clusters, and may be referred to as type III cavities, which may be nearly spherical cavities in reality, or type I or II cavities viewed nearly along the jet axis.

The remainder of the paper is organized as follows. Following a preliminary study of radial elongations of a sample of observed X-ray cavities in Sec. 2, we investigate the impact of line-of-sight projection on radial elongations in Sec. 3. By comparing with the results from a suite of hydrodynamic jet simulations, we then discuss what the observations of $\tau$ may reveal about the physics of mechanical AGN feedback in Sec. 4. We summarize our main results in Section \ref{section:discussion}.

\begin{deluxetable}{lcccc}
\label{tab1}
\tablenum{1}
\tablecaption{Properties of X-ray Cavities}
\tablewidth{0pt}
\tablehead{
 \colhead{System}&\colhead{$R_{\rm l}$}&\colhead{ $R_{\rm w}$}&\colhead{$d$} &\colhead{References} \\
 \colhead{}&\colhead{(kpc)}&\colhead{  (kpc)}&\colhead{ (kpc)} &\colhead{}
}
\startdata
A85   &6.3&8.9&21&1, 2\\
  A262   & 5.4&3.4&8.7&1, 3, 4\\ & 5.7&3.4&8.1& \\
   Perseus   &7.3&9.1&9.4&1, 5, 6\\ & 4.7&8.2&6.5&  \\ & 7.3&17&28&  \\ & 13&17&39&\\       
  2A 0335+096   & 6.5&9.3&23&1, 7\\ 
  A478   & 5.5&3.4&9&1, 8 \\  & 5.6&3.4&9&  \\
 MS 0735.6+7421   & 110&87&160&1, 9\\  & 130&89&180&  \\
4C+55.16  & 10&7.5&16&1, 10 \\   & 13&9.4&22&  \\
RBS 797  & 13&8.5&24&1, 11 \\   & 9.7&9.7&20&  \\
M84  & 1.6&1.6&2.3&1, 12 \\ & 2.1&1.2&2.5&\\
M87  & 2.3&1.4&2.8&1, 13 \\  & 1.6&0.8&2.2&  \\
Centaurus  & 2.4&3.3&6&1, 14 \\ & 1.6&3.3&3.5&  \\
HCG 62  & 4.3&5.0&8.4&1, 15 \\  & 4.0&4.0&8.6&  \\
Zw 2701& 8.75&12.25&18.9&16\\ &10.5&14.0&19.25&\\
 A3581 & 3.5&2.6&4.6& 1, 17\\  & 3.2&2.7&3.8& \\  & 3.8&8.4&24&17 \\
     MACS J1423.8+2404 & 9.4&9.4&16&1 \\  & 9.4&9.4&17&\\     
   A2052& 7.9&11&11&1, 18\\    & 6.2&6.5&6.7&\\   
 A2199& 12.1&8.5&23&19\\  & 14.7&9.9&23&\\  
 3C 388&15&15&27&1, 20 \\  & 10&24&21&\\
3C 401&12&12&15& 1 \\  & 12&12&15&\\
 Cygnus A&29&17&43&1, 21\\  & 34&23&45& \\
 A2597&7.1&7.1&23&1, 22 \\   & 10&7.1&23& \\
  A4059&20&10&23&1, 23 \\  & 9.2&9.2&19&\\
\enddata
 
\end{deluxetable}

 \begin{deluxetable}{lcccc}
\label{tab1}
\tablenum{1}
\tablecaption{Properties of X-ray Cavities (Continued)}
\tablewidth{0pt}
\tablehead{
 \colhead{System}&\colhead{$R_{\rm l}$}&\colhead{ $R_{\rm w}$}&\colhead{$d$} &\colhead{Refs} \\
 \colhead{}&\colhead{(kpc)}&\colhead{  (kpc)}&\colhead{ (kpc)} &\colhead{}
}
\startdata
 Hydra A  & 20.5&12.4&24.9&24\\   & 21&12.3&25.6&\\ & 31.5&47.2&100.8&\\ & 20.9&29&59.3& \\
    & 99.7&105&225.6&\\  & 50.1&67.7&104.3&  \\  
RX J1532.9+3021&14.4&17.3&28&25\\  & 12.1&14.9&39&\\ 
    NGC 5813&0.95&0.95&1.3&26\\& 1.03&0.93&1.4&\\   &3.9&3.9&7.7&  \\
   & 2.2&2.9&4.9&\\    &2.4&2.8&9.3&\\& 3.0&5.2&22.2&\\   &4.4&8.0&18&\\
\enddata
\vspace{0.3cm}
References --- (1) \citet{rafferty06}; (2) \citet{durret05}; (3) \citet{clarke09}; (4) \citet{blanton04}; (5) \citet{fabian02}; (6) \citet{fabian00}; (7) \citet{mazzotta03}; (8) \citet{sun03}; (9) \citet{mcnamara09}; (10) \citet{hl11}; (11) \citet{doria12}; (12) \citet{fj01}; (13) \citet{forman07}; (14) \citet{fabian05}; (15) \citet{gitti10}; (16) \citet{vagshette16}; (17) \citet{canning13}; (18) \citet{blanton11}; (19) \citet{nulsen13}; (20) \citet{kraft06}; (21) \citet{wilson06}; (22) \citet{clarke05}; (23) \citet{heinz02}; (24) \citet{wise07}; (25) \citet{hl13}; (26) \citet{randall11}
\end{deluxetable}

\section{Radial Elongations of Observed Cavities}

\label{section3}
 \begin{figure}
   \centering
     \epsscale{0.75}
\includegraphics[width=0.45\textwidth]{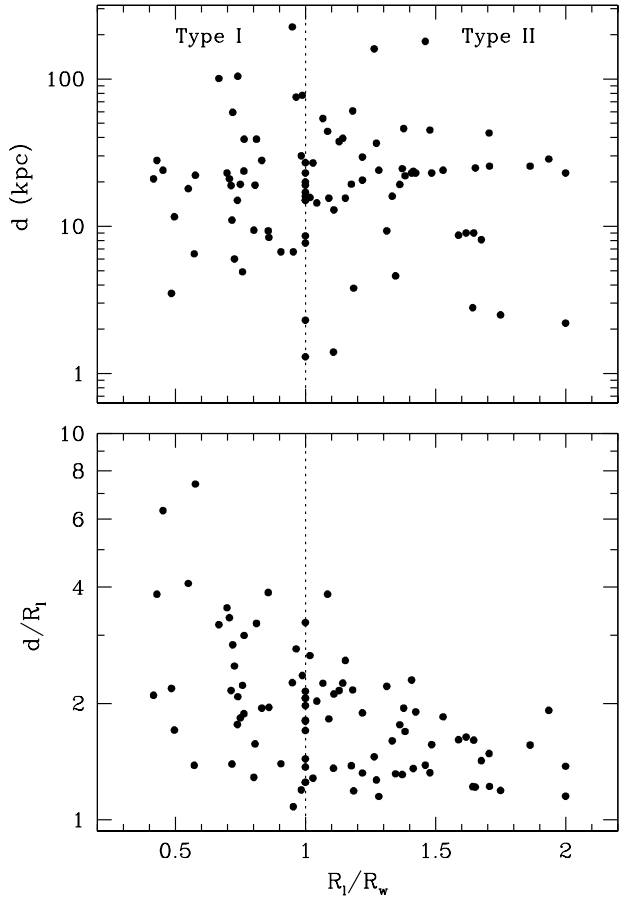} 
\caption{{\it Top panel}: The distance of the cavity center to the host system's center $d$ vs. the radial elongation $\tau=R_{\rm l}/R_{\rm w}$. No strong correlation exists between $d$ and $\tau$. {\it Bottom panel}: $d/R_{\rm l}$ vs. $\tau$. Old cavities have gone through buoyant evolution in the ICM, and are expected to have higher values of $d/R_{\rm l}$. There exists a general trend that the value of $\tau$ decreases as $d/R_{\rm l}$ increases. The vertical dotted line denotes $\tau=1$, which separates type-I cavities with $\tau<1$ and type-II cavities with $\tau>1$.}
 \label{plot2}
 \end{figure} 

 \begin{figure}
   \centering
     \epsscale{0.75}
\includegraphics[width=0.45\textwidth]{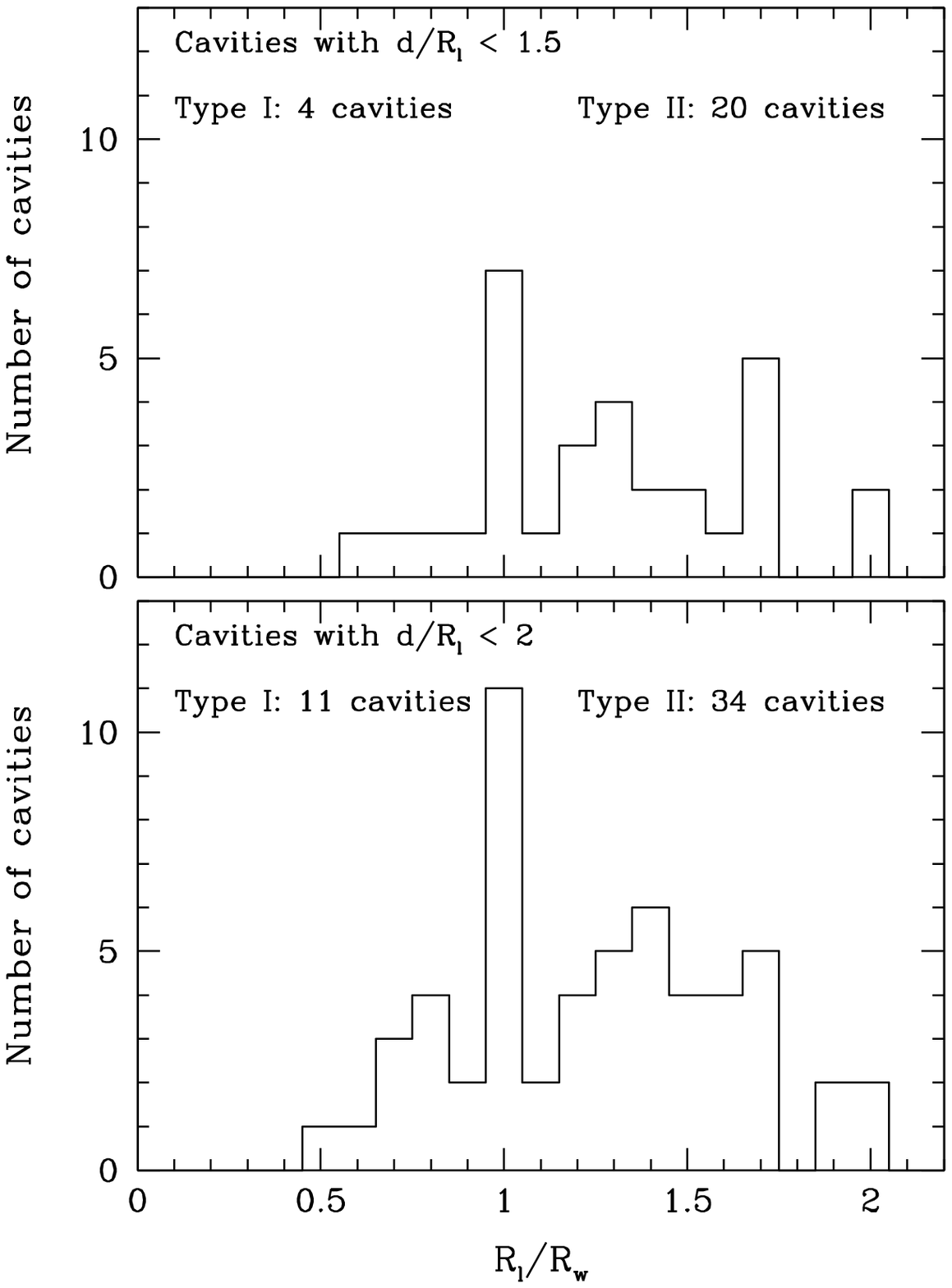} 
\caption{Histograms of the radial elongation $\tau=R_{\rm l}/R_{\rm w}$ in the relatively young X-ray cavities with $d/R_{\rm l}<1.5$ (top) and $d/R_{\rm l}<2$ (bottom). Bin widths are 0.1 in $\tau$. For young X-ray cavities, both type I and II cavities exist, and the latter dominates. The ratio between type I and II cavities increases from around $1:5$ for young cavities with $d/R_{\rm l}<1.5$ to around $1:3$ for those with $d/R_{\rm l}<2$.}
 \label{plot3}
 \end{figure} 
 
\subsection{The Cavity Sample}

In this section, we present a preliminary study of radial elongations in a sample of observed X-ray cavities drawn from the literature. Our cavity sample was mainly drawn from two large cavity samples in \citet{rafferty06} and \citet{hlavacek12}, respectively. 

The \citet{rafferty06} sample is a large sample of local and low-redshift X-ray cavities, most of which have also been used in other studies (e.g., \citealt{birzan04}; \citealt{diehl08}; \citealt{birzan08}; \citealt{birzan20}). \citet{rafferty06} provide the relevant parameters of each cavity in their sample, in particular, the distance of the cavity center to the host system's center (hereafter denoted as $d$), the semi-major and semi-minor axes. Some cavities have relatively low contrast with respect to their surroundings, and are assigned a value of ``3" for the figure of merit in \citet{rafferty06}. For accuracy, in our sample we exclude these poorly defined cavities without bright rims. We additionally determine the elongation of each cavity by eye in X-ray images published in the literature (as specifically listed in the rightmost column in Table \ref{tab1}). We find that all the cavities in our sample are elongated along either the jet direction or the angular direction perpendicular to the jet direction, and consequently we determine the values of $R_{\rm l}$ and $R_{\rm w}$ for each cavity according to the values of the semi-major and semi-minor axes given in \citet{rafferty06}. We updated the parameters ($R_{\rm l}$, $R_{\rm w}$, and $d$) of the cavities in Hydra A according to \citet{wise07}, those in Zw 2701 according to \citet{vagshette16}, and those in A2199 according to \citet{nulsen13}. We added one more cavity in A3581 observed by \citet{canning13}. We also supplemented our sample with two cavities in RX J1532.9+3021 \citep{hl13} and seven cavities in NGC 5813 \citep{randall11}. The parameters and references of these $60$ cavities are listed in Table \ref{tab1}.

The \citet{hlavacek12} sample includes 31 X-ray cavities in 20 galaxy clusters located at the redshift range of $0.3\leq z \leq 0.7$. The values of $R_{\rm l}$, $R_{\rm w}$, and $d$ of these cavities are explicitly listed in Table 3 of \citet{hlavacek12}. Combining with the cavities listed in Table 1, our cavity sample comprises a set of 91 cavities in 45 host systems, including 42 galaxy clusters, 2 galaxy groups (HCG 62 and NGC 5813) and one galaxy (M84). As located at relatively high redshifts, the cavities in the \citet{hlavacek12} sample usually do not have very high contrast with respect to their surroundings. However, we stress that our main results in the paper do not change qualitatively if we exclude the \citet{hlavacek12} sample from our analysis.

\subsection{Radial Elongations and Implications}

We first investigate the cavity elongation with respect to the jet direction in our sample. For each cavity listed in Table \ref{tab1}, we visually inspect its elongation in the X-ray images of its host system published in the corresponding references listed in the rightmost column in Table \ref{tab1}. We find that all the cavities listed in Table \ref{tab1} are either type-I cavities elongated along the angular direction ($R_{\rm l}<R_{\rm w}$), type-II cavities elongated along the jet direction ($R_{\rm l}>R_{\rm w}$), or nearly circular type-III cavities ($R_{\rm l}\approx R_{\rm w}$). For type-I cavities, the position angle between the cavity's semi-major axis and the jet direction is $90^{\circ}$. For type-II cavities, the position angle is $0^{\circ}$. This result strongly suggests that X-ray cavities are not subject to significant rotation during their evolution in the ICM. Otherwise, a large fraction of the cavities would be elongated along random directions with respect to the jet direction. It also implies that the observed difference between type I and II cavities in the cavity elongation with respect to the jet direction is not due to rotation in the ICM. Gas motions in the ICM may shift or bend X-ray cavities (e.g., \citealt{fabian00}; \citealt{venturi13}; \citealt{pm13}), but our result suggests that the kpc-scale rotation in the ICM velocity field is not significant. In other words, the level of turbulence in the inner regions of galaxy clusters may be relatively low, consistent with recent HITOMI observations of the Perseus cluster \citep{hitomi16}.

The distribution of the cavities in our sample is illustrated in Figure \ref{plot2}, which shows the diagrams of $d$ vs $\tau$ (top) and $d/R_{\rm l}$ vs $\tau$ (bottom). The value of radial elongation $\tau$ ranges between $0.4$ and $2$, and the centers of most cavities are located within $1\lesssim d \lesssim 100$ kpc from the host system's center. Figure  \ref{plot2} indicates that, while no clear correlation exists between $d$ and $\tau$, the value of $\tau$ roughly decreases as $d/R_{\rm l}$ increases. Cavities with high values of $d$ do not directly correspond to old cavities, as a big young cavity may be created directly with a large value of $d\sim R_{\rm l}$. Old cavities have gone through buoyant evolution in the ICM, and are instead expected to have high values of $d/R_{\rm l}$. Thus, the bottom panel of Figure 2 suggests that the value of $\tau$ decreases as a cavity rises buoyantly in the ICM. In other words, a cavity tends to become more elongated along the angular direction as it rises buoyantly in the ICM. An extreme case of this evolution is the pancake-shaped northwestern ghost cavity in Perseus \citep{fabian00,churazov01}, and this trend is also consistent with the predictions in hydrodynamic simulations (e.g., Figure 5 of \citealt{guo16}). Furthermore, the dearth of type-II cavities with $d/R_{\rm l}>2$ shown in the bottom panel of Figure 2 suggests that type-II cavities may evolve into type-I cavities as they rise buoyantly in the ICM.

While the shape of old cavities may change substantially during the buoyant evolution, the intrinsic shape of young cavities may be used to probe jet properties, as indicated by hydrodynamic simulations of \citet{guo15} and \citet{guo16}. Figure 3 shows the histograms of the radial elongation $\tau$ in the relatively young X-ray cavities with $d/R_{\rm l}<1.5$ (top) and $d/R_{\rm l}<2$ (bottom). It is clear that both type I and II young cavities exist, and the latter dominates. The ratio between type I and II cavities is around $1:5$ for young cavities with $d/R_{\rm l}<1.5$, and increases to around $1:3$ for cavities with $d/R_{\rm l}<2$, possibly due to buoyant evolution. As shown in Sec. 3, the inclination angle between the jet direction and the line of sight affects the observed value of $\tau$, but projection effect does not change type I cavities into type II cavities, or vice versa.

Remarkably, Figures 2 and 3 also indicate that there exist a large number of nearly circular type-III cavities with  $\tau \approx 1$. While some of them may be intrinsically spherical cavities, many of them may be type I or II cavities viewed along lines of sight close to the jet direction, as further discussed in Section 3.

 \begin{figure}
   \centering
     \epsscale{0.45}
\includegraphics[width=0.45\textwidth]{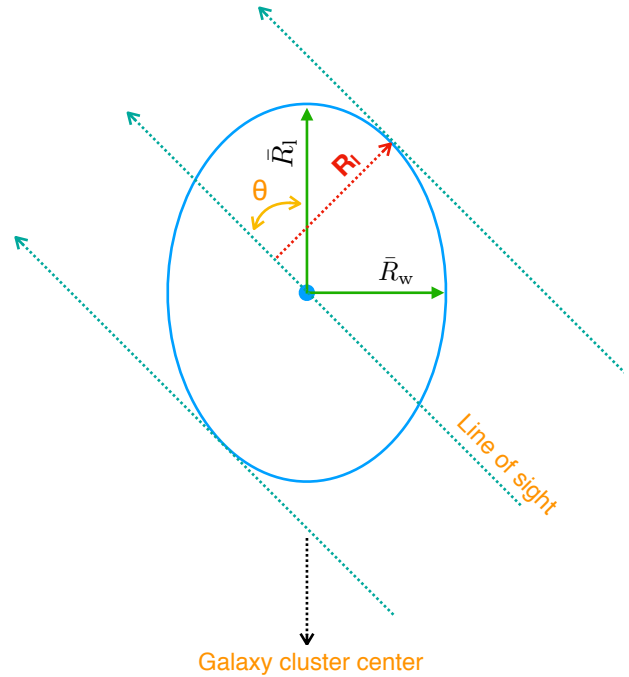} 
\caption{Sketch of parallel projection of an X-ray cavity onto the sky. The cavity is approximated as a spheroid with an intrinsic semi axis $\bar{R}_{l}$ along the jet direction, and two equal semi axes $\bar{R}_{\rm w}$ along two directions perpendicular to the jet direction. When projected onto the sky, ${R}_{l}$ is the apparent semi axis along the projected jet direction, while ${R}_{\rm w}=\bar{R}_{\rm w}$ is the semi axis along the angular direction perpendicular to the jet direction. $\theta$ is the inclination angle between the line of sight and the radial direction from the cluster center to the cavity center.}
 \label{plot4}
 \end{figure} 
 
  \begin{figure}
   \centering
     \epsscale{0.45}
\includegraphics[width=0.45\textwidth]{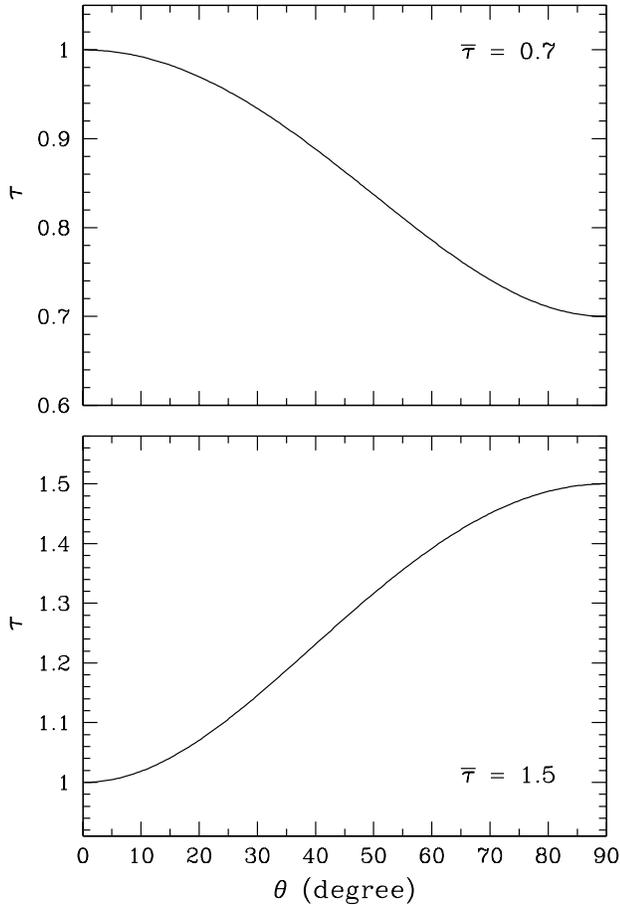} 
\caption{Impact of parallel projection along lines of sight on two idealized spheroidal cavities with intrinsic radial elongations $\bar{\tau}=0.7$ (top) and $1.5$ (bottom). $\theta$ is the inclination angle between the line of sight and the jet direction (see Fig. 4). It is clear that, depending on the inclination angle, parallel projection leads to an apparent radial elongation ranging from its intrinsic value $\tau=\bar{\tau}$ if $\theta=90^{\circ}$ to $\tau=1$ if $\theta=0^{\circ}$.} 
 \label{plot5}
 \end{figure} 

\section{Projection Effect on Radial Elongations}

Observed X-ray cavities are parallel projections of three-dimensional (3D) low-density ICM cavities along lines of sight onto the sky. Considering the evolution of an axisymmetric jet in a spherically-symmetric ICM, the created low-density cavity is expected to be axisymmetric around the jet axis. As a ``zeroth-order" approximation, the cavity may thus be approximated as a spheroid with a semi axis $\bar{R}_{l}$ along the jet direction, and two equal semi axes $\bar{R}_{\rm w}$ along two directions perpendicular to the jet direction. The cavity volume can be written as $V=4\pi \bar{R}_{l} \bar{R}_{\rm w}^{2}/3$, and the intrinsic radial elongation of this 3D cavity may be defined as $\bar{\tau}\equiv \bar{R}_{\rm l}/\bar{R}_{\rm w}$. In this section, we investigate how line-of-sight projections affect the observed value of radial elongation $\tau$.

Figure 4 shows a sketch of parallel projection of a 3D X-ray cavity onto the sky. For a 3D cavity with the intrinsic radial elongation $\bar{\tau}\equiv \bar{R}_{\rm l}/\bar{R}_{\rm w}$, the observed value of radial elongation $\tau=R_{\rm l}/R_{\rm w}$ depends on the inclination angle $\theta$ between the line of sight and the jet direction. When projected onto the sky, ${R}_{l}$ is the apparent semi axis along the projected jet direction on the sky, while ${R}_{\rm w}=\bar{R}_{\rm w}$ is the semi axis along the angular direction perpendicular to the projected jet direction. Here we use Figure 4 to facilitate the derivation of ${R}_{l}$, which may depend on $\bar{R}_{\rm l}$, $\bar{R}_{\rm w}$, and $\theta$. It is obvious that ${R}_{l}=\bar{R}_{\rm w}$ and $\tau=1$ if $\theta=0^{\circ}$, and ${R}_{l}=\bar{R}_{\rm l}$ and $\tau=\bar{\tau}$ if $\theta=90^{\circ}$.

Considering a Cartesian coordinate system $(x, z)$ with the origin at the cavity center, the line of sight passing through the origin can be written as $x=-z \text{~tan} \theta$, and the coordinates of an arbitrary point located at the cavity surface (ellipse) may be written as $(\bar{R}_{\rm w} \text{~cos} \phi, \bar{R}_{\rm l}\text{~sin} \phi)$. For the values of $\theta$ and $\phi$, we consider $0^{\circ} \leq \theta \leq 90^{\circ}$ and $0^{\circ} \leq \phi \leq 90^{\circ}$. The distance $l$ between a point $(\bar{R}_{\rm w} \text{~cos~} \phi, \bar{R}_{\rm l}\text{~sin~} \phi)$ at the cavity surface and a point $(-z \text{~tan~} \theta, z)$ on the line of sight passing through the cavity center can be written as
  \begin{eqnarray}
l^{2}= (\bar{R}_{\rm w}\text{~cos} \phi+z \text{~tan} \theta)^{2}+(\bar{R}_{\rm l}\text{~sin} \phi-z)^{2}{\rm .} \label{cdistance}
 \end{eqnarray}

The distance of the point $(\bar{R}_{\rm w} \text{~cos} \phi, \bar{R}_{\rm l}\text{~sin} \phi)$ to the line of sight $x=-z \text{~tan} \theta$ can then be evaluated according to Eq. (\ref{cdistance}) with the value of $z$ derived from the condition $\partial l^{2}/\partial z=0$, which gives $z=\bar{R}_{\rm l}\text{~sin} \phi \text{~cos}^{2}\theta-\bar{R}_{\rm w}\text{~cos} \phi \text{~sin}\theta \text{~cos}\theta$. Therefore one has
  \begin{eqnarray}
l= \bar{R}_{\rm w}\text{~cos} \phi \text{~cos}\theta+\bar{R}_{\rm l}\text{~sin} \phi \text{~sin} \theta    {\rm .} \label{cdistance2}
 \end{eqnarray}
As illustrated in Figure 4, the apparent semi axis along the projected jet direction on the sky ($R_{\rm l}$) corresponds to the maximum value of $l$ in Eq. (\ref{cdistance2}), which occurs at the value of $\phi$ determined by the condition $\partial l/\partial \phi=0$, i.e.,
  \begin{eqnarray}
\text{tan} \phi=\bar{\tau}\text{tan} \theta {\rm ~,}
\label{cdistance3}
 \end{eqnarray}
where $\bar{\tau}= \bar{R}_{\rm l}/\bar{R}_{\rm w}$. As ${R}_{\rm w}=\bar{R}_{\rm w}$, Eq. (\ref{cdistance2}) can be rewritten as
  \begin{eqnarray}
\tau= &\text{~cos} \phi \text{~cos}\theta+\bar{\tau}\text{~sin} \phi \text{~sin} \theta \nonumber \\  
=&\sqrt{\text{~cos}^{2}\theta +\bar{\tau}^{2}\text{~sin}^{2} \theta} {\rm ,} \label{cdistance4}
 \end{eqnarray}
where we have used the value of $\phi$ in Eq. (\ref{cdistance3}). Due to the axisymmetry around the jet axis, Eq. (\ref{cdistance4}) also holds for $90^{\circ}\leq \theta \leq 180^{\circ}$. For the special case of spherical cavities with $\bar{\tau}=1$, Eqs. (\ref{cdistance3}) and (\ref{cdistance4}) reduce to $\phi=\theta$ and $\tau=1$, respectively.

Eq. (\ref{cdistance4}) indicates that, depending on the inclination angle $\theta$, parallel projection leads to an apparent radial elongation $\tau$ ranging from its intrinsic value $\tau=\bar{\tau}$ if $\theta=90^{\circ}$ to $\tau=1$ if $\theta=0^{\circ}$, as clearly illustrated in Figure 5 for a type-I cavity with $\bar{\tau}=0.7$ (top) and a type-II cavity with $\bar{\tau}=1.5$ (bottom). While projection effect affects the observed value of $\tau$ by making cavities appear more circular, it does not change type I cavities into type II cavities, or vice versa. Thus the quantity ratio between type I and II young cavities investigated in Sec. 2 is not affected by projection effect. We also note that due to the measurement uncertainties in $R_{\rm l}$ and $R_{\rm w}$, a small but substantial fraction of type-I and -II cavities with small values of $|\theta|$ (i.e., viewed along lines of sight close to the jet axis) may be classified as type-III cavities with $\tau \approx 1$. 

Eq. (\ref{cdistance4}) can also be used to improve the measurement of the cavity volume $V=4\pi R_{\rm l} R_{\rm w}^{2}/3$, which is often adopted to estimate the energetics and power of mechanical AGN feedback (e.g., \citealt{hlavacek12}). Combining ${R}_{\rm w}=\bar{R}_{\rm w}$ with Eq. (\ref{cdistance4}), one can easily derive the correlation between $R_{\rm l}$ and $\bar{R}_{\rm l}$: $R_{\rm l}^{2}=\bar{R}_{\rm w}^{2}\text{~cos}^{2}\theta +\bar{R}_{\rm l}^{2}\text{~sin}^{2} \theta$.
For type-I cavities, $\bar{R}_{\rm w}>\bar{R}_{\rm l}$ and thus $R_{\rm l}>\bar{R}_{\rm l}$. Therefore $V=4\pi R_{\rm l} R_{\rm w}^{2}/3$ is an upper limit of the real cavity volume $\bar{V}=4\pi \bar{R}_{\rm l} \bar{R}_{\rm w}^{2}/3$. Similarly, for type-II cavities, $R_{\rm l}<\bar{R}_{\rm l}$ and $V=4\pi R_{\rm l} R_{\rm w}^{2}/3$ is a lower limit of the real cavity volume.

If there is a very large sample of young X-ray cavities detected in the future, the distribution of $\theta$ may be considered to be random, and Eq. (\ref{cdistance4}) may then be used to derive the intrinsic probability distribution function (PDF) of $\bar{\tau}$ from the observed PDF of $\tau$. For type-I cavities, one has
  \begin{eqnarray}
f(\tau)d\tau & \propto (\int^{\tau}_{0} g(\bar{\tau})\left|\frac{d\theta(\tau,\bar{\tau})}{d\tau}\right|d\bar{\tau})d\tau \nonumber \\
&=\frac{\tau}{\sqrt{1-\tau^{2}}}(\int^{\tau}_{0} \frac{g(\bar{\tau})}{\sqrt{\tau^{2}-\bar{\tau}^{2}}}d\bar{\tau})d\tau  {\rm ~,} 
 \end{eqnarray}
where $|d\theta(\tau,\bar{\tau})/d\tau|=\tau/\sqrt{(1-\tau^{2})(\tau^{2}-\bar{\tau}^{2})}$ is derived from Eq. (\ref{cdistance4}), and $f(\tau)d\tau$ and $g(\bar{\tau})d\bar{\tau}$ are the PDFs of $\tau$ and $\bar{\tau}$ of type-I cavities, respectively.  Similarly, for type-II cavities, one has
  \begin{eqnarray}
f(\tau)d\tau  \propto \frac{\tau}{\sqrt{\tau^{2}-1}}(\int^{\infty}_{\tau} \frac{g(\bar{\tau})}{\sqrt{\bar{\tau}^{2}-\tau^{2}}}d\bar{\tau})d\tau  {\rm .} 
 \end{eqnarray}

It should be noted that the derivations in this Section rely on the assumption that X-ray cavities are spheroidal and axisymmetric around the jet axis. For triaxial cavities, the calculations on the projection effect would be much more complex, and the results may differ substantially from our current results. However, large X-ray cavities are produced by powerful jets with very small cross sections, and it may be natural to assume that they are roughly axisymmetric around the original jet axis. In this case, the main results in this Section, e.g, projection effect does not change type I cavities into type II cavities, may still hold qualitatively, although detailed calculations are required for confirmation.

\section{Intrinsic Radial Elongations in Simulations}
\label{section2}

 \begin{figure}
   \centering
\includegraphics[width=0.45\textwidth]{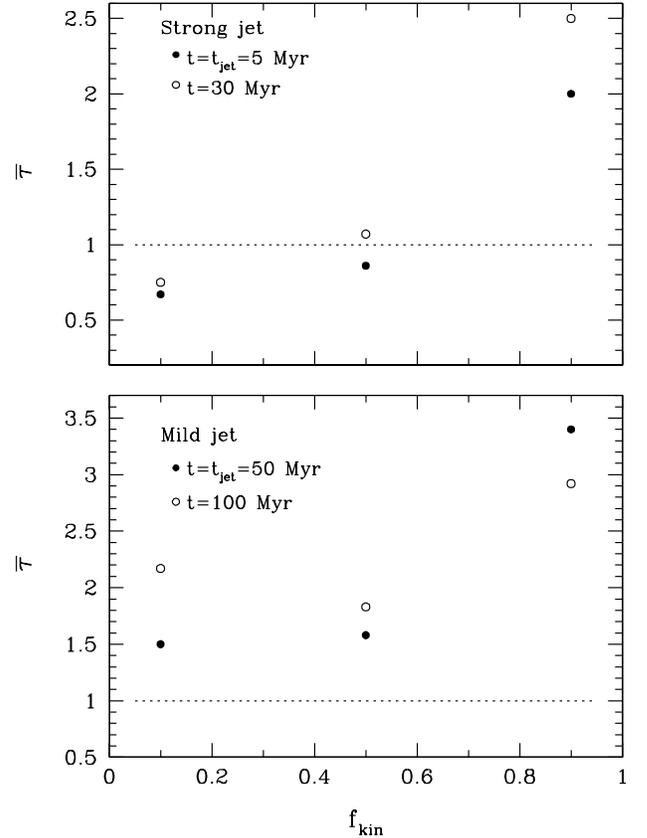} 
\caption{Intrinsic radial elongations $\bar{\tau}$ of young X-ray cavities in a series of hydrodynamic jet simulations of \citet{duan20} as a function of the jet kinetic fraction $f_{\rm kin}$. These simulated X-ray cavities are produced by AGN jets in a realistic galaxy cluster, and the jets in all the simulations presented here have the same jet energy $E_{\rm inj}=2.3\times 10^{60}$ erg. The top panel shows the results of three simulations of strong jets with a short jet duration $t_{\rm inj}=5$ Myr at two times $t=5$ Myr and $30$ Myr, while the bottom panel shows those of three simulations of mild jets with a long jet duration $t_{\rm inj}=50$ Myr at two times $t=50$ Myr and $100$ Myr. The cavities presented here are relatively young, still attached to or just slightly detached from the cluster center. The simulations presented in the top and bottom panels are illustrated in Figures 4 and 5 of \citet{duan20}, respectively. The horizontal dotted line in each panel refers to type-III cavities with $\bar{\tau}=1$.}
 \label{plot6}
 \end{figure} 

In this section, we investigate what the observations of radial elongations of young X-ray cavities may tell us about the physics of mechanical AGN feedback. In our previous studies \citep{guo15, guo16}, we found that in a typical smooth spherically-symmetric ICM, the shape of young cavities is mainly affected by the properties of AGN jets. Here we adopt our recent hydrodynamic jet simulations in \citet{duan20} to summarize the dependence of the intrinsic radial elongation $\bar{\tau}$ of young X-ray cavities on jet properties, particularly its kinetic fraction. In these simulations, we use thermal jets carrying both thermal and kinetic energies, which are initialized at the jet base on kpc scales (more specifically at $z_{\rm inj}=1$ kpc from the cluster center). The kinetic fraction $f_{\rm kin}=1-f_{\rm th}$ is defined as the ratio of the kinetic energy density to the total energy density within the jet at its base. Here $f_{\rm th}$ is the jet's thermal fraction at the jet base. For more details of the setup and results of these simulations, we refer the reader to \citet[see also \citealt{guo18} and \citealt{duan18}]{duan20}.

Figure \ref{plot6} shows $\bar{\tau}$ of young cavities as a function of $f_{\rm kin}$ in a series of strong (top) and mild (bottom) jet simulations in \citet{duan20}. These simulated X-ray cavities are produced by AGN jets in a realistic galaxy cluster (Abell 1795), and the jets all have the same total energy $E_{\rm inj}=2.3\times 10^{60}$ erg. For a given total energy $E_{\rm inj}$, \citet{duan20} demonstrate that there exists a characteristic radius $R_{\rm fb}$ within which the total ICM thermal energy equals $E_{\rm inj}$, which defines a characteristic jet power $P_{\rm fb}=E_{\rm inj}/t_{\rm s}$, where $t_{\rm s}$ is the sound crossing time across $R_{\rm fb}$. For a given $E_{\rm inj}$, a jet with power much higher (lower) than $P_{\rm fb}$ may be considered to be a strong (mild) jet, and shock dissipation in the ICM is much more significant in strong AGN outbursts than in mild ones \citep{duan20}. While the jet radius and velocity are fixed in these simulations, the value of $f_{\rm kin}$ varies as we adopt different values of the internal jet density and thermal energy density at the jet base. The strong and mild jet simulations used to make Figure \ref{plot6} are illustrated in Figures 4 and 5 of \citet{duan20}, respectively. 

For strong jet outbursts, the value of $\bar{\tau}$ of young cavities increases with $f_{\rm kin}$, and as clearly shown in the top panel of Figure \ref{plot6}, the transition from type I to type II cavities occurs roughly at $f_{\rm kin}\sim 0.5$. In other words, thermal-energy-dominated jets tend to produce type-I cavities, while kinetic-energy-dominated jets tend to produce type-II cavities. Compared to strong jets, mild jets with the same total energy have a higher jet duration, producing young cavities that are more elongated along the jet direction and have higher values of $\bar{\tau}$, as clearly seen in Figure \ref{plot6} (see also relevant discussions in Sec. 3.1 of \citealt{guo15}). While $\bar{\tau}$ also roughly increases with $f_{\rm kin}$ for mild jet outbursts, both thermal-energy-dominated and kinetic-energy-dominated mild jets produce type II cavities with $\bar{\tau}>1$.

Now we can use our simulation results to interpret the observations of $\tau$ of young X-ray cavities presented in Sec. 2. In our cavity sample, both type-I and -II young cavities exist, and type II young cavities dominate. As shown in Sec. 3, while projection effect affects the observed value of $\tau$, it does not change type I cavities into type II cavities, or vice versa. Therefore, the existence of type-I young cavities indicates that some AGN jets are strong jets energetically dominated by thermal energy on kpc scales, consistent with \citet{guo16} using the morphology (top-wideness) of young X-ray cavities. These jets are not dominated by the kinetic energy on kpc scales as often assumed in the literature (e.g., \citealt{gaspari11}; \citealt{yang16}; \citealt{guo18}; \citealt{martizzi19}; \citealt{bambic19}), and may alternatively be energetically dominated by cosmic rays as recently suggested (e.g., \citealt{guo08a}; \citealt{guo11}; \citealt{ruszkowski17}; \citealt{yang19}; \citealt{wang20}). It has also been demonstrated by \citet{duan20} that strong non-kinetic-energy-dominated jets are much more effective in delaying the onset of cooling catastrophe than kinetic-energy-dominated jets with the same power.
 
However, the dominance of type-II over type-I young cavities suggests that these strong jets dominated by non-kinetic energies are not dominant in mechanical AGN feedback. If AGN jets are dominated by strong jets with relatively short durations, most of them should be energetically dominated by the kinetic energy. However, if most AGN jets are energetically dominated by non-kinetic energies, our results suggest that they must be mainly mild jets with relatively long durations. In the jet interior, the ratio between the kinetic energy density $e_{\rm kin}$ and the thermal energy density $e_{\rm th}$ is
 \begin{eqnarray}
\frac{e_{\rm kin}}{e_{\rm th}}=\frac{\gamma(\gamma-1)}{2}M_{\rm int}^{2}{\rm ~,}  
\end{eqnarray}
where $M_{\rm int}$ is the internal Mach number defined as the ratio of the jet speed to the sound speed in the jet interior \citep{guo16}. Therefore, kinetic-energy dominated jets are internally supersonic jets, while non-kinetic-energy dominated jets correspond to internally subsonic jets.

\section{Summary and Discussion}
\label{section:discussion}

Mechanical AGN feedback is usually thought to play a key role in the evolution of massive galaxies, galaxy groups and clusters. However, the particle and energy content of AGN jets that mediate this feedback process is still far from clear. {\it Chandra} and {\it XMM-Newton} observations have detected a large number of X-ray cavities, apparently evolved from mechanical AGN feedback in the hot gaseous halo of these systems and potentially containing important information on the physics of mechanical AGN feedback. The enthalpy of X-ray cavities has already been extensively used to estimate the energetics and power of mechanical AGN feedback. Here we present a preliminary study of radial elongations of a large sample of X-ray cavities drawn from the literature, and investigate the implications on the physics of mechanical AGN feedback.

All the 91 X-ray cavities in our sample are type-I cavities elongated along the angular direction (perpendicular to the jet direction), type-II cavities elongated along the jet direction, or nearly circular type-III cavities. This suggests that X-ray cavities are not subject to significant rotation during their evolution in the ICM, and implies that the observed difference in radial elongation between type I and II cavities is not due to the cavity rotation in the ICM. Our result also suggests that the kpc-scale rotation in the ICM velocity field is not significant and the level of turbulence in the inner 100-kpc regions of galaxy clusters may be relatively low, consistent with recent HITOMI observations of the Perseus cluster \citep{hitomi16}.

The value of radial elongation $\tau$ may vary as a cavity rises buoyantly in the ICM. We explored this issue with the potential correlations between $\tau$ and $d$, and between $\tau$ and $d/R_{\rm l}$. Here $d$ is the distance of the cavity center to the host system's center. All the three types of cavities are seen in our sample with different value of $d$ varying between 1 and 100 kpc, and there is not a clear trend between $\tau$ and $d$. In contrast, a trend between $\tau$ and $d/R_{\rm l}$ exists in our sample, and $\tau$ roughly decreases with the increasing value of $d/R_{\rm l}$. This suggests that $d/R_{\rm l}$ may be a better indicator determining whether a cavity has gone through significant buoyant evolution in the ICM, and X-ray cavities tend to become more elongated along the angular direction (with lower values of $\tau$) as they rise buoyantly in the ICM. This is consistent with the predictions in hydrodynamic simulations. The dearth of type-II cavities with $d/R_{\rm l}>2$ suggests that type-II cavities may evolve into type-I cavities as they rise buoyantly in the ICM.
 
Young X-ray cavities have not gone through significant buoyant evolution, and their radial elongations may tell us about the properties of AGN jets that created them. In our cavity sample, both type I and II young cavities exist, and the latter dominates. The observed value of $\tau$ is expected to be affected by projection effect. Assuming that X-ray cavities are spheroidal and axisymmetric around the jet axis, we derive an analytical relation between the intrinsic radial elongation $\bar{\tau}$ and the observed value of $\tau$: $\tau= \sqrt{\text{~cos}^{2}\theta +\bar{\tau}^{2}\text{~sin}^{2} \theta}$, which depends on the inclination angle $\theta$. The value of $\tau$ generally lies between 1 and $\bar{\tau}$, indicating that projection effect makes cavities appear more circular, but does not change type-I cavities into type-II ones, or vice versa. The relation can be used to improve the measurement of the cavity volume and confirms that some type-III cavities may be intrinsically type-I or -II cavities viewed along lines of sight close to the jet axis.

We investigate the intrinsic radial elongations of young cavities in a suite of hydrodynamic jet simulations, and find that $\bar{\tau}$ typically increases with the kinetic fraction of AGN jets. As demonstrated in \citet{duan20}, for a given jet energy, there exists a characteristic jet power that separates short-duration strong jets and long-duration mild jets. Irrespective of the kinetic fraction, mild jets always produce type-II cavities. However, for strong jets, thermal-energy-dominated jets tend to produce type-I cavities, while kinetic-energy-dominated jets produce type-II cavities. The existence of type-I young cavities indicates that some AGN jets are strong and dominated by non-kinetic energies on kpc scales, such as thermal energy or cosmic rays. However, the dominance of type-II over type-I young cavities suggests that these jets are not dominant in mechanical AGN feedback. If most jets are dominated by non-kinetic energies, they should be mainly long-duration mild jets. However, if most jets are strong, they must be mainly dominated by the kinetic energy. Additional observations are required to further infer if most jets are strong jets dominated by the kinetic energy or mild jets dominated by non-kinetic energies. 

Radio observations of mechanical AGN feedback indicate that there is a dichotomy between Fanaroff-Riley (FR) type I and II radio sources \citep{fr74}. While both FR I and II radio sources exist in galaxy clusters, the former dominates. According to the values of radial elongations derived from X-ray observations, our study suggests that both type I and II young X-ray cavities exist, and the latter dominates. While the archetypical FR II radio jets in Cygnus A produce type-II young cavities, FR II radio sources are not very common in galaxy clusters, suggesting that many FR I radio sources also produce type-II young X-ray cavities. It is thus possible that some FR I radio sources produce type-I cavities, while others produce type-II cavities. The difference may be caused by the duration of AGN jets, as short-duration (long-duration) jets tend to produce type I (II) cavities. As radial elongation is also affected by the jet's kinetic fraction, the difference may also be caused by the properties and the dissipation/entrainment history of AGN jets on sub-kpc scales.

\section*{Acknowledgments}

The author would like to thank Zhen-Ya Zheng and Minfeng Gu for helpful discussions, and the referee, Chris Reynolds, for valuable comments that improved the presentation of the paper. This work was supported partially by National Natural Science Foundation of China (Grant No. 11873072 and 11633006), Natural Science Foundation of Shanghai (No. 18ZR1447100), and Chinese Academy of Sciences through the Key Research Program of Frontier Sciences (No. QYZDB-SSW-SYS033 and QYZDJ-SSW-SYS008). The simulations analyzed in this work were performed using the high performance computing resources in the Core Facility for Advanced Research Computing at Shanghai Astronomical Observatory.

\bibliography{ms}

\begin{thebibliography}{}
\expandafter\ifx\csname natexlab\endcsname\relax\def\natexlab#1{#1}\fi
\providecommand{\url}[1]{\href{#1}{#1}}

\bibitem[{{Bambic} \& {Reynolds}(2019)}]{bambic19}
{Bambic}, C.~J., \& {Reynolds}, C.~S. 2019, \apj, 886, 78

\bibitem[{{B{\^i}rzan} {et~al.}(2008){B{\^i}rzan}, {McNamara}, {Nulsen},
  {Carilli}, \& {Wise}}]{birzan08}
{B{\^i}rzan}, L., {McNamara}, B.~R., {Nulsen}, P.~E.~J., {Carilli}, C.~L., \&
  {Wise}, M.~W. 2008, \apj, 686, 859

\bibitem[{{B{\^i}rzan} {et~al.}(2004){B{\^i}rzan}, {Rafferty}, {McNamara},
  {Wise}, \& {Nulsen}}]{birzan04}
{B{\^i}rzan}, L., {Rafferty}, D.~A., {McNamara}, B.~R., {Wise}, M.~W., \&
  {Nulsen}, P.~E.~J. 2004, \apj, 607, 800

\bibitem[{{B{\^\i}rzan} {et~al.}(2020){B{\^\i}rzan}, {Rafferty}, {Br{\"u}ggen},
  {Botteon}, {Brunetti}, {Cuciti}, {Edge}, {Morganti}, {R{\"o}ttgering}, \&
  {Shimwell}}]{birzan20}
{B{\^\i}rzan}, L., {Rafferty}, D.~A., {Br{\"u}ggen}, M., {et~al.} 2020, \mnras,
  496, 2613

\bibitem[{{Blanton} {et~al.}(2011){Blanton}, {Randall}, {Clarke}, {Sarazin},
  {McNamara}, {Douglass}, \& {McDonald}}]{blanton11}
{Blanton}, E.~L., {Randall}, S.~W., {Clarke}, T.~E., {et~al.} 2011, \apj, 737,
  99

\bibitem[{{Blanton} {et~al.}(2004){Blanton}, {Sarazin}, {McNamara}, \&
  {Clarke}}]{blanton04}
{Blanton}, E.~L., {Sarazin}, C.~L., {McNamara}, B.~R., \& {Clarke}, T.~E. 2004,
  \apj, 612, 817

\bibitem[{{Boehringer} {et~al.}(1993){Boehringer}, {Voges}, {Fabian}, {Edge},
  \& {Neumann}}]{boehringer93}
{Boehringer}, H., {Voges}, W., {Fabian}, A.~C., {Edge}, A.~C., \& {Neumann},
  D.~M. 1993, \mnras, 264, L25

\bibitem[{{Canning} {et~al.}(2013){Canning}, {Sun}, {Sanders}, {Clarke},
  {Fabian}, {Giacintucci}, {Lal}, {Werner}, {Allen}, {Donahue}, {Edge},
  {Johnstone}, {Nulsen}, {Salom{\'e}}, \& {Sarazin}}]{canning13}
{Canning}, R.~E.~A., {Sun}, M., {Sanders}, J.~S., {et~al.} 2013, \mnras, 435,
  1108

\bibitem[{{Churazov} {et~al.}(2001){Churazov}, {Br{\"u}ggen}, {Kaiser},
  {B{\"o}hringer}, \& {Forman}}]{churazov01}
{Churazov}, E., {Br{\"u}ggen}, M., {Kaiser}, C.~R., {B{\"o}hringer}, H., \&
  {Forman}, W. 2001, \apj, 554, 261

\bibitem[{{Clarke} {et~al.}(2009){Clarke}, {Blanton}, {Sarazin}, {Anderson},
  {Gopal-Krishna}, {Douglass}, \& {Kassim}}]{clarke09}
{Clarke}, T.~E., {Blanton}, E.~L., {Sarazin}, C.~L., {et~al.} 2009, \apj, 697,
  1481

\bibitem[{{Clarke} {et~al.}(2005){Clarke}, {Sarazin}, {Blanton}, {Neumann}, \&
  {Kassim}}]{clarke05}
{Clarke}, T.~E., {Sarazin}, C.~L., {Blanton}, E.~L., {Neumann}, D.~M., \&
  {Kassim}, N.~E. 2005, \apj, 625, 748

\bibitem[{{Croston} \& {Hardcastle}(2014)}]{croston14}
{Croston}, J.~H., \& {Hardcastle}, M.~J. 2014, \mnras, 438, 3310

\bibitem[{{Croston} {et~al.}(2008){Croston}, {Hardcastle}, {Birkinshaw},
  {Worrall}, \& {Laing}}]{croston08}
{Croston}, J.~H., {Hardcastle}, M.~J., {Birkinshaw}, M., {Worrall}, D.~M., \&
  {Laing}, R.~A. 2008, \mnras, 386, 1709

\bibitem[{{Croston} {et~al.}(2011){Croston}, {Hardcastle}, {Mingo}, {Evans},
  {Dicken}, {Morganti}, \& {Tadhunter}}]{croston11}
{Croston}, J.~H., {Hardcastle}, M.~J., {Mingo}, B., {et~al.} 2011, \apjl, 734,
  L28

\bibitem[{{Diehl} {et~al.}(2008){Diehl}, {Li}, {Fryer}, \&
  {Rafferty}}]{diehl08}
{Diehl}, S., {Li}, H., {Fryer}, C.~L., \& {Rafferty}, D. 2008, \apj, 687, 173

\bibitem[{{Doria} {et~al.}(2012){Doria}, {Gitti}, {Ettori}, {Brighenti},
  {Nulsen}, \& {McNamara}}]{doria12}
{Doria}, A., {Gitti}, M., {Ettori}, S., {et~al.} 2012, \apj, 753, 47

\bibitem[{{Duan} \& {Guo}(2018)}]{duan18}
{Duan}, X., \& {Guo}, F. 2018, \apj, 861, 106

\bibitem[{{Duan} \& {Guo}(2020)}]{duan20}
---. 2020, \apj, 896, 114

\bibitem[{{Durret} {et~al.}(2005){Durret}, {Lima Neto}, \& {Forman}}]{durret05}
{Durret}, F., {Lima Neto}, G.~B., \& {Forman}, W. 2005, \aap, 432, 809

\bibitem[{{Fabian} {et~al.}(2002){Fabian}, {Celotti}, {Blundell}, {Kassim}, \&
  {Perley}}]{fabian02}
{Fabian}, A.~C., {Celotti}, A., {Blundell}, K.~M., {Kassim}, N.~E., \&
  {Perley}, R.~A. 2002, \mnras, 331, 369

\bibitem[{{Fabian} {et~al.}(2005){Fabian}, {Sanders}, {Taylor}, \&
  {Allen}}]{fabian05}
{Fabian}, A.~C., {Sanders}, J.~S., {Taylor}, G.~B., \& {Allen}, S.~W. 2005,
  \mnras, 360, L20

\bibitem[{{Fabian} {et~al.}(2000){Fabian}, {Sanders}, {Ettori}, {Taylor},
  {Allen}, {Crawford}, {Iwasawa}, {Johnstone}, \& {Ogle}}]{fabian00}
{Fabian}, A.~C., {Sanders}, J.~S., {Ettori}, S., {et~al.} 2000, \mnras, 318,
  L65

\bibitem[{{Fanaroff} \& {Riley}(1974)}]{fr74}
{Fanaroff}, B.~L., \& {Riley}, J.~M. 1974, \mnras, 167, 31P

\bibitem[{{Finoguenov} \& {Jones}(2001)}]{fj01}
{Finoguenov}, A., \& {Jones}, C. 2001, \apjl, 547, L107

\bibitem[{{Forman} {et~al.}(2007){Forman}, {Jones}, {Churazov}, {Markevitch},
  {Nulsen}, {Vikhlinin}, {Begelman}, {B{\"o}hringer}, {Eilek}, {Heinz},
  {Kraft}, {Owen}, \& {Pahre}}]{forman07}
{Forman}, W., {Jones}, C., {Churazov}, E., {et~al.} 2007, \apj, 665, 1057

\bibitem[{{Gaspari} {et~al.}(2011){Gaspari}, {Melioli}, {Brighenti}, \&
  {D'Ercole}}]{gaspari11}
{Gaspari}, M., {Melioli}, C., {Brighenti}, F., \& {D'Ercole}, A. 2011, \mnras,
  411, 349

\bibitem[{{Gitti} {et~al.}(2010){Gitti}, {O'Sullivan}, {Giacintucci}, {David},
  {Vrtilek}, {Raychaudhury}, \& {Nulsen}}]{gitti10}
{Gitti}, M., {O'Sullivan}, E., {Giacintucci}, S., {et~al.} 2010, \apj, 714, 758

\bibitem[{{Guo}(2015)}]{guo15}
{Guo}, F. 2015, \apj, 803, 48

\bibitem[{{Guo}(2016)}]{guo16}
---. 2016, \apj, 826, 17

\bibitem[{{Guo} {et~al.}(2018){Guo}, {Duan}, \& {Yuan}}]{guo18}
{Guo}, F., {Duan}, X., \& {Yuan}, Y.-F. 2018, \mnras, 473, 1332

\bibitem[{{Guo} \& {Mathews}(2011)}]{guo11}
{Guo}, F., \& {Mathews}, W.~G. 2011, \apj, 728, 121

\bibitem[{{Guo} \& {Oh}(2008)}]{guo08a}
{Guo}, F., \& {Oh}, S.~P. 2008, \mnras, 384, 251

\bibitem[{{Heinz} {et~al.}(2002){Heinz}, {Choi}, {Reynolds}, \&
  {Begelman}}]{heinz02}
{Heinz}, S., {Choi}, Y.-Y., {Reynolds}, C.~S., \& {Begelman}, M.~C. 2002,
  \apjl, 569, L79

\bibitem[{{Hitomi Collaboration} {et~al.}(2016){Hitomi Collaboration},
  {Aharonian}, {Akamatsu}, {Akimoto}, {Allen}, {Anabuki}, {Angelini}, {Arnaud},
  {Audard}, {Awaki}, {Axelsson}, {Bamba}, {Bautz}, {Blandford}, {Brenneman},
  {Brown}, {Bulbul}, {Cackett}, {Chernyakova}, {Chiao}, {Coppi}, {Costantini},
  {de Plaa}, {den Herder}, {Done}, {Dotani}, {Ebisawa}, {Eckart}, {Enoto},
  {Ezoe}, {Fabian}, {Ferrigno}, {Foster}, {Fujimoto}, {Fukazawa}, {Furuzawa},
  {Galeazzi}, {Gallo}, {Gandhi}, {Giustini}, {Goldwurm}, {Gu}, {Guainazzi},
  {Haba}, {Hagino}, {Hamaguchi}, {Harrus}, {Hatsukade}, {Hayashi}, {Hayashi},
  {Hayashida}, {Hiraga}, {Hornschemeier}, {Hoshino}, {Hughes}, {Iizuka},
  {Inoue}, {Inoue}, {Ishibashi}, {Ishida}, {Ishikawa}, {Ishisaki}, {Itoh},
  {Iyomoto}, {Kaastra}, {Kallman}, {Kamae}, {Kara}, {Kataoka}, {Katsuda},
  {Katsuta}, {Kawaharada}, {Kawai}, {Kelley}, {Khangulyan}, {Kilbourne},
  {King}, {Kitaguchi}, {Kitamoto}, {Kitayama}, {Kohmura}, {Kokubun}, {Koyama},
  {Koyama}, {Kretschmar}, {Krimm}, {Kubota}, {Kunieda}, {Laurent}, {Lebrun},
  {Lee}, {Leutenegger}, {Limousin}, {Loewenstein}, {Long}, {Lumb}, {Madejski},
  {Maeda}, {Maier}, {Makishima}, {Markevitch}, {Matsumoto}, {Matsushita},
  {McCammon}, {McNamara}, {Mehdipour}, {Miller}, {Miller}, {Mineshige},
  {Mitsuda}, {Mitsuishi}, {Miyazawa}, {Mizuno}, {Mori}, {Mori}, {Moseley},
  {Mukai}, {Murakami}, {Murakami}, {Mushotzky}, {Nagino}, {Nakagawa},
  {Nakajima}, {Nakamori}, {Nakano}, {Nakashima}, {Nakazawa}, {Nobukawa},
  {Noda}, {Nomachi}, {O'Dell}, {Odaka}, {Ohashi}, {Ohno}, {Okajima}, {Ota},
  {Ozaki}, {Paerels}, {Paltani}, {Parmar}, {Petre}, {Pinto}, {Pohl}, {Porter},
  {Pottschmidt}, {Ramsey}, {Reynolds}, {Russell}, {Safi-Harb}, {Saito},
  {Sakai}, {Sameshima}, {Sato}, {Sato}, {Sato}, {Sawada}, {Schartel},
  {Serlemitsos}, {Seta}, {Shidatsu}, {Simionescu}, {Smith}, {Soong}, {Stawarz},
  {Sugawara}, {Sugita}, {Szymkowiak}, {Tajima}, {Takahashi}, {Takahashi},
  {Takeda}, {Takei}, {Tamagawa}, {Tamura}, {Tamura}, {Tanaka}, {Tanaka},
  {Tanaka}, {Tashiro}, {Tawara}, {Terada}, {Terashima}, {Tombesi}, {Tomida},
  {Tsuboi}, {Tsujimoto}, {Tsunemi}, {Tsuru}, {Uchida}, {Uchiyama}, {Uchiyama},
  {Ueda}, {Ueda}, {Ueno}, {Uno}, {Urry}, {Ursino}, {de Vries}, {Watanabe},
  {Werner}, {Wik}, {Wilkins}, {Williams}, {Yamada}, {Yamaguchi}, {Yamaoka},
  {Yamasaki}, {Yamauchi}, {Yamauchi}, {Yaqoob}, {Yatsu}, {Yonetoku}, {Yoshida},
  {Yuasa}, {Zhuravleva}, \& {Zoghbi}}]{hitomi16}
{Hitomi Collaboration}, {Aharonian}, F., {Akamatsu}, H., {et~al.} 2016, \nat,
  535, 117

\bibitem[{{Hlavacek-Larrondo} {et~al.}(2012){Hlavacek-Larrondo}, {Fabian},
  {Edge}, {Ebeling}, {Sanders}, {Hogan}, \& {Taylor}}]{hlavacek12}
{Hlavacek-Larrondo}, J., {Fabian}, A.~C., {Edge}, A.~C., {et~al.} 2012, \mnras,
  421, 1360

\bibitem[{{Hlavacek-Larrondo} {et~al.}(2011){Hlavacek-Larrondo}, {Fabian},
  {Sanders}, \& {Taylor}}]{hl11}
{Hlavacek-Larrondo}, J., {Fabian}, A.~C., {Sanders}, J.~S., \& {Taylor}, G.~B.
  2011, \mnras, 415, 3520

\bibitem[{{Hlavacek-Larrondo} {et~al.}(2013){Hlavacek-Larrondo}, {Allen},
  {Taylor}, {Fabian}, {Canning}, {Werner}, {Sand ers}, {Grimes}, {Ehlert}, \&
  {von der Linden}}]{hl13}
{Hlavacek-Larrondo}, J., {Allen}, S.~W., {Taylor}, G.~B., {et~al.} 2013, \apj,
  777, 163

\bibitem[{{Hlavacek-Larrondo} {et~al.}(2015){Hlavacek-Larrondo}, {McDonald},
  {Benson}, {Forman}, {Allen}, {Bleem}, {Ashby}, {Bocquet}, {Brodwin},
  {Dietrich}, {Jones}, {Liu}, {Reichardt}, {Saliwanchik}, {Saro}, {Schrabback},
  {Song}, {Stalder}, {Vikhlinin}, \& {Zenteno}}]{hlavacek15}
{Hlavacek-Larrondo}, J., {McDonald}, M., {Benson}, B.~A., {et~al.} 2015, \apj,
  805, 35

\bibitem[{{Kraft} {et~al.}(2006){Kraft}, {Azcona}, {Forman}, {Hardcastle},
  {Jones}, \& {Murray}}]{kraft06}
{Kraft}, R.~P., {Azcona}, J., {Forman}, W.~R., {et~al.} 2006, \apj, 639, 753

\bibitem[{{Li} {et~al.}(2015){Li}, {Bryan}, {Ruszkowski}, {Voit}, {O'Shea}, \&
  {Donahue}}]{li15}
{Li}, Y., {Bryan}, G.~L., {Ruszkowski}, M., {et~al.} 2015, \apj, 811, 73

\bibitem[{{Martizzi} {et~al.}(2019){Martizzi}, {Quataert},
  {Faucher-Gigu{\`e}re}, \& {Fielding}}]{martizzi19}
{Martizzi}, D., {Quataert}, E., {Faucher-Gigu{\`e}re}, C.-A., \& {Fielding}, D.
  2019, \mnras, 483, 2465

\bibitem[{{Mazzotta} {et~al.}(2003){Mazzotta}, {Edge}, \&
  {Markevitch}}]{mazzotta03}
{Mazzotta}, P., {Edge}, A.~C., \& {Markevitch}, M. 2003, \apj, 596, 190

\bibitem[{{McNamara} {et~al.}(2009){McNamara}, {Kazemzadeh}, {Rafferty},
  {B{\^i}rzan}, {Nulsen}, {Kirkpatrick}, \& {Wise}}]{mcnamara09}
{McNamara}, B.~R., {Kazemzadeh}, F., {Rafferty}, D.~A., {et~al.} 2009, \apj,
  698, 594

\bibitem[{{McNamara} \& {Nulsen}(2007)}]{mcnamara07}
{McNamara}, B.~R., \& {Nulsen}, P.~E.~J. 2007, \araa, 45, 117

\bibitem[{{McNamara} \& {Nulsen}(2012)}]{mcnamara12}
---. 2012, New Journal of Physics, 14, 055023

\bibitem[{{Nulsen} {et~al.}(2013){Nulsen}, {Li}, {Forman}, {Kraft}, {Lal},
  {Jones}, {Zhuravleva}, {Churazov}, {Sanders}, {Fabian}, {Johnson}, \&
  {Murray}}]{nulsen13}
{Nulsen}, P. E.~J., {Li}, Z., {Forman}, W.~R., {et~al.} 2013, \apj, 775, 117

\bibitem[{{Paterno-Mahler} {et~al.}(2013){Paterno-Mahler}, {Blanton},
  {Randall}, \& {Clarke}}]{pm13}
{Paterno-Mahler}, R., {Blanton}, E.~L., {Randall}, S.~W., \& {Clarke}, T.~E.
  2013, \apj, 773, 114

\bibitem[{{Rafferty} {et~al.}(2006){Rafferty}, {McNamara}, {Nulsen}, \&
  {Wise}}]{rafferty06}
{Rafferty}, D.~A., {McNamara}, B.~R., {Nulsen}, P.~E.~J., \& {Wise}, M.~W.
  2006, \apj, 652, 216

\bibitem[{{Randall} {et~al.}(2011){Randall}, {Forman}, {Giacintucci}, {Nulsen},
  {Sun}, {Jones}, {Churazov}, {David}, {Kraft}, {Donahue}, {Blanton},
  {Simionescu}, \& {Werner}}]{randall11}
{Randall}, S.~W., {Forman}, W.~R., {Giacintucci}, S., {et~al.} 2011, \apj, 726,
  86

\bibitem[{{Ruszkowski} {et~al.}(2017){Ruszkowski}, {Yang}, \&
  {Reynolds}}]{ruszkowski17}
{Ruszkowski}, M., {Yang}, H. Y.~K., \& {Reynolds}, C.~S. 2017, \apj, 844, 13

\bibitem[{{Soker}(2016)}]{soker16}
{Soker}, N. 2016, \nar, 75, 1

\bibitem[{{Sun} {et~al.}(2003){Sun}, {Jones}, {Murray}, {Allen}, {Fabian}, \&
  {Edge}}]{sun03}
{Sun}, M., {Jones}, C., {Murray}, S.~S., {et~al.} 2003, \apj, 587, 619

\bibitem[{{Vagshette} {et~al.}(2019){Vagshette}, {Naik}, \&
  {Patil}}]{vagshette19}
{Vagshette}, N.~D., {Naik}, S., \& {Patil}, M.~K. 2019, \mnras, 485, 1981

\bibitem[{{Vagshette} {et~al.}(2016){Vagshette}, {Sonkamble}, {Naik}, \&
  {Patil}}]{vagshette16}
{Vagshette}, N.~D., {Sonkamble}, S.~S., {Naik}, S., \& {Patil}, M.~K. 2016,
  \mnras, 461, 1885

\bibitem[{{Venturi} {et~al.}(2013){Venturi}, {Rossetti}, {Bardelli},
  {Giacintucci}, {Dallacasa}, {Cornacchia}, \& {Kantharia}}]{venturi13}
{Venturi}, T., {Rossetti}, M., {Bardelli}, S., {et~al.} 2013, \aap, 558, A146

\bibitem[{{Wang} {et~al.}(2020){Wang}, {Ruszkowski}, \& {Yang}}]{wang20}
{Wang}, C., {Ruszkowski}, M., \& {Yang}, H. Y.~K. 2020, \mnras, 493, 4065

\bibitem[{{Werner} {et~al.}(2019){Werner}, {McNamara}, {Churazov}, \&
  {Scannapieco}}]{werner19}
{Werner}, N., {McNamara}, B.~R., {Churazov}, E., \& {Scannapieco}, E. 2019,
  \ssr, 215, 5

\bibitem[{{Wilson} {et~al.}(2006){Wilson}, {Smith}, \& {Young}}]{wilson06}
{Wilson}, A.~S., {Smith}, D.~A., \& {Young}, A.~J. 2006, \apjl, 644, L9

\bibitem[{{Wise} {et~al.}(2007){Wise}, {McNamara}, {Nulsen}, {Houck}, \&
  {David}}]{wise07}
{Wise}, M.~W., {McNamara}, B.~R., {Nulsen}, P.~E.~J., {Houck}, J.~C., \&
  {David}, L.~P. 2007, \apj, 659, 1153

\bibitem[{{Yang} {et~al.}(2019){Yang}, {Gaspari}, \& {Marlow}}]{yang19}
{Yang}, H. Y.~K., {Gaspari}, M., \& {Marlow}, C. 2019, \apj, 871, 6

\bibitem[{{Yang} \& {Reynolds}(2016)}]{yang16}
{Yang}, H. Y.~K., \& {Reynolds}, C.~S. 2016, \apj, 818, 181

\end{thebibliography}

\label{lastpage}

\end{document}